\documentclass[aps,prb,floatfix,twocolumn]{revtex4}
\usepackage{epsfig}
\begin{document}

\hsize\textwidth\columnwidth\hsize\csname@twocolumnfalse\endcsname

\title{Optical orientation in bipolar spintronic devices}

\author{Jaroslav Fabian}
\affiliation{Institute for Theoretical Physics, University of
Regensburg, 93040 Regensburg, Germany}

\author{Igor \v{Z}uti\'{c}}
\address{Department of Physics, State University of New York at
Buffalo, Buffalo NY, 14260, USA}


\vskip1.5truecm
\begin{abstract}
Optical orientation is a highly efficient tool for the generation of
nonequilibrium spin polarization in semiconductors. Combined with
spin-polarized transport it offers new functionalities for
conventional electronic devices, such as pn junction bipolar diodes
or transistors. In nominally nonmagnetic junctions optical
orientation can provide a source for spin capacitance---the
bias-dependent nonequilibrium spin accumulation---or for
spin-polarized current in bipolar spin-polarized solar cells. In
magnetic junctions, the nonequilibrium spin polarization generated
by spin orientation in a proximity of an equilibrium magnetization
gives rise to the spin-voltaic effect (a realization of the
Silsbee-Johnson coupling), enabling efficient control of electrical
properties such as the I-V characteristics of the junctions by
magnetic and optical fields. This article reviews the main results
of investigations of spin-polarized and magnetic pn junctions, from
spin capacitance to the spin-voltaic effect.
\end{abstract}
\maketitle

\begin{figure}
\label{fig:1}
\end{figure}

\section{Introduction}

Semiconductor spintronics ~\cite{Fabian2007:APS, Zutic2004:RMP} has
inspired fascinating developments in spin-based electronics as well
as in our understanding of fundamental spin-related processes in
solids. In particular, we have learned how to control spin by
electric and optical means, and how the presence of spin is
manifested in electronic and optical~\cite{meier:1984} processes.
This article discusses selected device schemes inspired by optical
orientation of nonequilibrium spin, as well as by the fundamental
concept of spin-charge coupling, originally proposed by Silsbee and
Johnson for metals. ~\cite{Silsbee1980:BMR, Johnson1985:PRL}

The device settings we consider are spin-polarized pn junctions and
magnetic diodes, forming building blocks of bipolar spintronics.
While we review the most important results and present selected
functionalities of the devices, we first briefly comment on the
methodology we use to study them. Our proposals are based on two
approaches: (i) fully numerical self-consistent calculations of the
spin-polarized drift-diffusion transport and (ii) analytical
modeling in the regime of low injection.

The main idea behind the numerical approach is designing a
discretization scheme to solve a set of coupled differential
equations for drift-diffusion transport. We typically consider
electrons as spin-polarized, leaving holes unpolarized. This is
justified by the fast spin relaxation of holes in zinc-blend
semiconductors which could be used as the materials of choice in
bipolar spintronic devices (silicon alone is ineffective as far as
optical orientation is concerned). While there is no technical
difficulty of involving spin-polarized holes, the presentation is
focused on the most important effects when considering only
electrons as the carriers of spin.

Let us denote by $D_n$ the electron diffusion coefficient, by $n =
n_\uparrow + n_\downarrow$ and $s=n_\uparrow - n_\downarrow$ the
electron and electron spin densities, by $\phi_t$ the electrostatic
potential inside the device (consisting of the equilibrium built-in
field as well as the field coming from the external emf source; the
label $t$ stands for total), and by $2\zeta$ the spin-splitting, in
general spatially varying, of the conduction band. The
drift-diffusion prescription for the electron particle and electron
spin currents is
\begin{eqnarray}
J_n = J_{n\uparrow} + J_{n\downarrow} & = & D_n (n\phi'_t + s\zeta' - n'), \\
J_s = J_{n\uparrow} - J_{n\downarrow} & = & D_n (s\phi_t' + n\zeta'
- s').
\end{eqnarray}
We express the energy ($\phi_t$ and $\zeta$) in the units of thermal
energy, $k_B T$. We also assume the semiconductors are
nondegenerate, with all the donors/acceptors fully polarized. In
fact, for the numerical illustrations we consider room temperature
(and give bias in volts). A similar expression as for $J_n$ can be
written for the hole current, $J_p$, for holes in the valence band.

Electrons and holes are coupled by the recombination processes.
Since we consider optically active zinc-blende materials, the
presumed coupling is through the electron-hole recombination, of the
rate coefficient $r$, described by the following continuity
equations:
\begin{eqnarray}
J_n' & = & -r (np - n_0p_0) + G, \\
J_s' & = & -r (s p - s_0 p_0) - \frac{\delta \tilde{s}}{\tau_s} +
G^s.
\end{eqnarray}
The hole density is $p$ and the equilibrium densities (at absent
bias and optical generation) are indexed with zero. The continuity
equation for spin takes into account that the spin density decays by
both the electron-hole recombination as well as by intrinsic spin
relaxation, of the rate of the inverse of the spin relaxation time
$\tau_s$. Important, $\delta \tilde{s}$ is the deviation of the
instantaneous spin density from the spin density that would be the
equilibrium one for the instantaneous electron density: $ \delta
\tilde{s} = s - P_0n$, where $P$ is the equilibrium spin
polarization. The spin decays to $\tilde{s}_0= P_0n$, not to $s_0 =
P_0 n_0$, as $\tau_s$ reflects intrinsic spin relaxation processes
\cite{meier:1984}; electron-hole recombination, which also degrades
spin, is considered explicitly. Optical generation of charges or
spin can be described according to the specific situation, either
through boundary conditions on the nonequilibrium spin or charge, if
the light is incident through the edges, or through the constant
photoexcitation rates $G$ for electrons and $G^s$ for the spin. In
zinc-blend semiconductors $G^s=G/2$, since the spin polarization of
electrons at the time of pair creation is $50\%$. \cite{meier:1984}
In addition to the external spin sources due to spin injection or
optical orientation, there will be spin polarization in our bipolar
system due to spin-orbit coupling. Such intrinsic spin polarizations
will be typically much smaller (say, on the order of 0.1\%) than
those considered here. \cite{Tse2005:PRB}.

Finally, the self-consistent loop is completed by coupling the
transport to the local charge density via the Poisson equation,
\begin{equation}
\phi_t''=-\rho(q/\epsilon k_B T),
\end{equation}
with $\rho$ collecting all the 
contributions (generally, spatially dependent)
consisting of ionized donors ($N_d$), acceptors ($N_a$), and the
electron ($n$) and hole ($p$) densities:
\begin{equation}
\rho = q (N_d - N_a + p -n).
\end{equation}
We stress that the effects of the electric field are included by
default in this scheme.

While the numerical approach is essential in modeling general
transport conditions, at low injections (small biases---typically
applied voltages smaller than the energy band gap of the host
material) the numerical simulations can be remarkably well
substituted by a few assumptions allowing for rather simple
analytical modeling. Following the original insight of Shockley
~\cite{Shockley:1950}, who has pioneered the low injection transport
modeling of bipolar devices, we have formulated the assumption
governing spin-polarized drift diffusion in generic devices
containing spin sources as well as the spin splitting of the carrier
energy bands.~\cite{Fabian2002:PRB} The conditions, valid in the low
injection limit, are: (a) the bulk regions (those away from the
depletion layers) are neutral, (b) there is a thermal
quasiequilibrium across the depletion layer at applied biases and at
the presence of source spin (this essentially means that
spin-resolved chemical potentials are constant across the depletion
layer), and (c) the spin current is continuous across the depletion
layer. The last condition can be relaxed and substituted by a
current continuity model of spin relaxation in the depletion layer.

The analytical theory and its ramifications for spin-polarized
bipolar transport is worked out in Ref. \onlinecite{Fabian2002:PRB}.
Here we list the most important conclusions, which follow from the
generalized Shockley conditions a) and b) above. First, there is a
nice relation connecting the spin density polarizations,
\begin{equation}\label{eq:polarization}
P=\frac{s}{n},
\end{equation}
across the depletion layer. If the p region is on the left ($L$) of
the layer, and the n region on the right ($R$), the relation can be
stated as,
\begin{equation} \label{eq:PL}
P_L = \frac{P_{0L} (1-P^2_{0R}) + \delta P_R (1 -
P_{0L}P_{0R})}{1-P^2_{0R} + \delta P_R (P_{0L} - P_{0R})}.
\end{equation}
The zero-indexed polarizations are the equilibrium ones. In
addition,
\begin{equation}
\delta P = P -P_0,
\end{equation}
denotes the nonequilibrium part of the polarization. We will find
this equation useful in discussing the physics of spin injection
across the depletion layer in spin-polarized pn junctions.

There are also relations connecting the carrier and spin densities
across the depletion layer. We have,
\begin{eqnarray} \label{eq:nL}
n_L & = & n_{0L} e^V \left (1 + \delta P_R \frac{P_{0L}-P_{0R}}{1-P_{0R}^2}  \right ), \\
s_L & = & s_{0L} e^V \left (1 + \frac{\delta P_R}{P_{0L}} \frac{1 -
P_{0L}P_{0R}}{1-P_{0R}^2}  \right );
\end{eqnarray}
recall that the applied bias $V$ is measured in the units of thermal
energy $k_B T$. For spin-unpolarized case, $s_{0L} = \delta P_R=0$,
the voltage dependence of $n_L$, the electron density in the p
region close to the depletion layer follows the  typical
rectification behavior (actually giving rise to the rectification
I-V characteristic); $n_L$ is small for reverse biases ($V<0$),
while it exponentially increases for forward biases ($V>0$) with
increasing $V$. In the spin-polarized case the electron density
$n_L$ and the resulting electron current are modified by $\delta
P_R$, the amount of nonequilibrium spin in the n region, but {\it
only} if there is an equilibrium spin polarization present. This is
the statement of the spin-charge coupling, applied to pn junctions.
Compared to metal physics, the coupling is actually exponentially
amplified by the applied forward bias, leading to a strong
giant-magnetoresistance like behavior of magnetic pn junction
diodes, as will be discussed later.

The relations Eqs. \ref{eq:PL} and \ref{eq:nL} contain an unknown
parameter, $\delta P_R$. Typically we set our boundary conditions
far from the depletion layer, simulating carrier and spin injection.
The nonequilibrium spin at the depletion layer is a result of a
self-consistent spin distribution throughout the junction, and to
obtain it the last condition, c), needs to be invoked. The result is
an analytical formula for $\delta  P_R$ which allows fully
analytical modeling of the spin injection
physics.~\cite{Fabian2002:PRB} We do not reproduce this formula
here.

In addition to the usual definition of spin polarization, given in
Eq. \ref{eq:polarization}, we also introduce the spin polarization
of the electric current,
\begin{equation}
P_J = \frac{J_s}{J} = \frac{J_s}{J_n + J_p},
\end{equation}
as the ratio of the spin and (total) charge current. Since unlike
the charge current the spin current is not conserved (and is in
general not uniform throughout the space), the spin current
polarization $P_J$ is typically space dependent. Normally we are
concerned about the current spin polarization at the contacts with
electrodes, as the spin injection efficiency is directly
proportional to it; see the discussion of spin injection in Refs.
\onlinecite{Zutic2004:RMP} and \onlinecite{Fabian2007:APS}.

Other bipolar devices, not discussed here, include spin
light-emitting ~\cite{Fiederling1999:N, Fiederling2003:APL,
Ohno1999:N, Young2002:APL, Jonker2000:PRB, Hanbicki2002:APLa,
Jiang2003:PRL, Zega2006:PRL} and, in some sense, spin Esaki diodes.
~\cite{Kohda2001:JJAP, Johnston2002:PRB} We also list alternative
spintronic device schemes that may be practical in the long run:
resonant tunneling diodes ~\cite{Ertler2006b:APL, Ertler2006a:APL,
Ertler2007a:PRB, Slobodskyy2003:PRL, Petukhov2002:PRL,
Ertler2008:PRL}, a room temperature spin-transference device
~\cite{Dery2006:PRB} or a scheme for reconfigurable logic
\cite{Dery2007:N} , unipolar magnetic diodes ~\cite{Flatte2001:APL},
to name a few.

Optical orientation can also be useful to detect spin in otherwise
optically inactive materials (such as silicon), by surrounding the
material with an optically active material (such as GaAs) and allow
for spin transfer. Such a scheme was proposed to detect spin
injection into silicon ~\cite{Zutic2006:PRL}, as a recently
experimentally demonstrated ~\cite{Jonker2007:NP, Erve2007:APL}
alternative way to an all-electron detection ~\cite{Appelbaum2007:N,
Huang2007:PRL, Zutic2007:N} in this prominent material.

Yet other important applications of optical spin orientation are
spin-polarized bipolar semiconductor lasers. It was demonstrated
that pumping a vertical-cavity surface emitting laser (VCSEL) with
optically generated spin-polarized carriers can lead to the lasing
operation at the reduced threshold current as compared to the
spin-unpolarized VCSEL.~\cite{Rudolph2003:APL, Rudolph2005:APL} For
a fixed pump power, the emitted power of the laser can be increased
by 400 \% by simply changing the degree of spin polarization of
pumped carriers.~\cite{Rudolph2003:APL} Both the threshold current
reduction and the change of emitted power with the variation of spin
polarization of pumped carriers have also been demonstrated in
electrically pumped VCSEL.~\cite{Holub2007:PRL,Holub2007:JPD} It was
recently shown~\cite{Gothgen2008:P} that spin VCSELs could exceed
previously accepted theoretical limit (50 \%) for the threshold
reduction and act as efficient nonlinear filters of
circularly-polarized light. While the work on spin VCSELs has
emerged only in the last several years, it already shows a clear
path towards practical realization. Revealing novel effects in
semiconductor spin lasers will closely depend on our ability to
better understand the interplay of optical orientation and bipolar
transport.

\section{Spin-polarized pn junctions: spin capacitance}

When nonequilibrium spin is injected in nominally nonmagnetic pn
junctions, we speak of spin-polarized pn junctions. The spin
injection can be done optically or electrically and, while the
specifics of the device depends on the injection mode---through
specific boundary conditions for the current or carrier
concentrations---for the principle of operation it does not matter.
In what follows we consider optical orientation which gives both
nonequilibrium spins and charges at the spin injection point; in the
case of an electrical spin injection carriers would be at thermal
equilibrium, in ohmic contacts.

The pn junction has three regions: p-region, n-region, and the
depletion layer in between. The operation of the junction depends on
in which region the nonequilibrium or source spin is injected. In
the following we consider only electrons as spin polarized; holes
are kept unpolarized. This is partly justified by the fact that
holes in bulk zinc-blende semiconductors lose spin polarization very
fast, on the order of the momentum relaxation time. If the spin is
injected into the p-region, the minority carrier (electrons) are
spin polarized, carrying the spin polarization through the depletion
layer into the n-region. The consequences are spin pumping as well
as the spin capacitance effect, as discussed below. If the spin is
injected into the depletion layer, or in a close proximity, the pn
junction functions as a spin-polarized solar cell, the subject of
the following section. Finally, if the spin is injected into the
n-region, orienting the majority carriers, ordinary spin injection
into the p-region follows under forward bias.

\begin{figure}
\centerline{\psfig{file=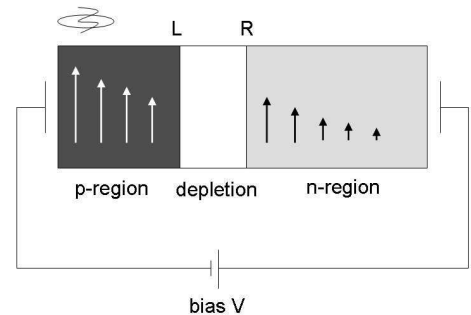,width=0.8\linewidth}}
\caption{Source spin is injected in the p-region of a nonmagnetic pn
junction. Spin is both accumulated in the p-region and pumped into
the n-region. Both effects depend on the applied bias $V$.}
\label{fig:p-injection}
\end{figure}

Suppose a nonequilibrium spin is injected into the p-region, as
illustrated in Fig. \ref{fig:p-injection}. The spin propagates by
drift-diffusion towards the depletion layer. Assuming the spin
injection point is within the spin diffusion length from the
depletion layer, a significant spin polarization survives there.
What happens to the spin in the depletion layer? The large built-in
field (typically tens of kV/cm) sweeps all the electrons to the
n-region. Since the spin is attached to the carriers, the spin
density follows. In the n-region the spin accumulates, creating a
spin density gradient and setting up a back-diffusion. In the steady
state process the spin influx balances the back-diffusion. We termed
this process spin pumping by the minority
channel.~\cite{Zutic2001:PRB} Interestingly, the injected spin
density is (spatially) amplified as it goes through the depletion
layer. On the other hand, the spin polarization is preserved, at
least at low biases; it is not amplified. This follows from a
general result, that the nonequilibrium spin polarizations at the
p-side ($\delta P_L$ for the left) and at the n-side ($\delta P_R$
for the right) of the depletion layer of a nonmagnetic pn junction
are the same, in the low injection limit (see Eq. \ref{eq:PL}):
\begin{equation} \label{eq:depltaPL}
\delta P_L = P_L= \delta P_R = P_R,
\end{equation}
as $P_{0L}=P_{0R}=0$ for nonmagnetic junctions. The argument, based
on generalized Shockley's conditions, is presented in Ref.
\onlinecite{Fabian2002:PRB}.

\begin{figure}
\centerline{\psfig{file=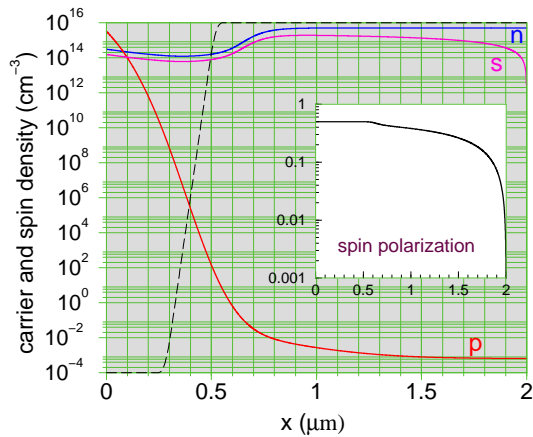,width=0.8\linewidth}}
\caption{Calculated profile for the electron ($n$), hole ($p$), and
electron spin ($s$) densities for a generic pn junction. The dashed
line indicates the doping density profile (scale not given)---the
p-region starts at the left, the depletion layer appears around 0.5
$\mu$m, followed by the n-region: $N_d(x)-N_a(x)$; the acceptor
(donor) density deep in the p (n) region is $N_a=3\times 10^{15}$
($N_d=5\times 10^{15}$) per cm$^3$. The inset displays the spin
polarization profile. Optical orientation occurs at $x=0$; the
electron-hole and spin generation are modeled by the boundary
conditions for $n$ and $s$, whose values can be read off of the
graph. There is no bulk illumination ($G=0$). From Ref.
\onlinecite{Zutic2001:PRB}.} \label{fig:density}
\end{figure}

In Fig. \ref{fig:density} we plot the results of a self-consistent
drift-diffusion calculation for a GaAs inspired nondegenerate pn
junction with optical spin injection (source spin) in the p region.
As optically both spin and electron-hole pairs are created, the
electron density in the p-region is also enhanced. The spin density
decays somewhat towards the depletion layer, essentially following
the decrease of the nonequilibrium electron density due to
electron-hole recombination processes. Both densities are strongly
enhanced in the n region, although the spin polarization itself
stays unchanged, disappearing at the far edge of the n region only.
The figure demonstrates both the spin pumping through the minority
channel as well as the spin density amplification.

\begin{figure}
\centerline{\psfig{file=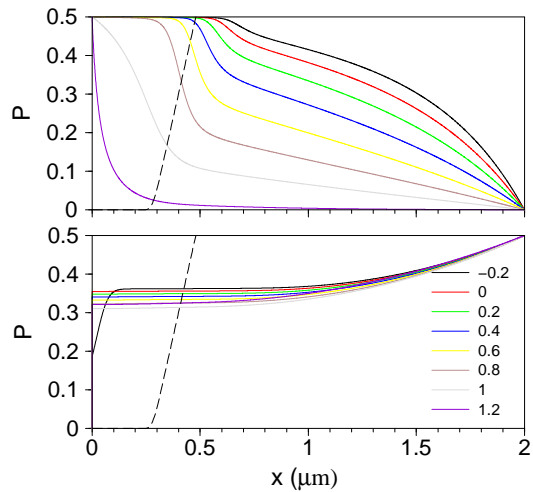,width=0.8\linewidth}}
\caption{Calculated spin polarization profiles for the pn junction
in Fig. \ref{fig:density}, for different biases $V$. The upper graph
shows the polarization for the source spin injection in the p
region, while the lower graph is for the source spin injection into
the n region. The dashed line indicates the doping profile, $N_d -
N_a$, defined in Fig. \ref{fig:density}. From Ref.
\onlinecite{Zutic2001:PRB}.} \label{fig:polarization}
\end{figure}

If we look at the spin polarization, plotted in Fig.
\ref{fig:polarization} we see at first a surprising decrease of its
magnitude with increasing the bias voltage. Recall that a negative
bias results in a reverse regime, at which only very little (so
called generation) current flows through the junction. In this
regime the depletion layer widens and the built-in electric field
increases. The intrinsic barrier for the carriers to cross from one
region to the other increases, inhibiting the current flow. For the
spin injection the reverse region means that the minority electrons
have reduced distances to go to the wider depletion layer (the spin
injection point is fixed), so that the electrons have less time to
recombine with holes and large electron and the accompanying spin
densities arrive at the boundary of the depletion layer. These
electrons are then swept by the increased built-in field into the n
region, pumping the spin there.

If the source spin is added to the n region, under forward biases
the spin can be injected into the p region. See Fig.
\ref{fig:polarization}. The spin injection process is
straightforward: spin is injected along with the nonequilibrium
electron density into the p region, in which both the spin and the
carrier density decay mainly by electron-hole recombination. In the
reverse bias the spin injection is severely inhibited.

\begin{figure}
\centerline{\psfig{file=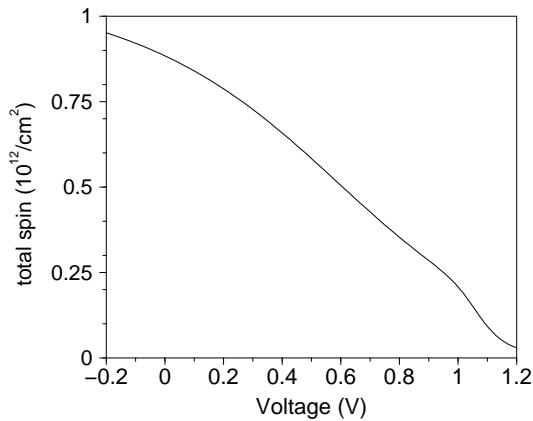,width=0.8\linewidth}} \caption{Spin
capacitance effect. The total (integrated) spin in the pn junction
due to the source spin injection in the p region. The total spin
decays steadily with increasing the bias, from the reverse (negative
$V$) to the forward (positive $V$) bias region. From Ref.
\onlinecite{Zutic2001:PRB}.} \label{fig:capacity}
\end{figure}

Spin-polarized pn junctions exhibit the  spin capacitance
phenomenon. Following an analogy with conventional charge
capacitance we define the spin capacitance as the change of the {\it
nonequilibrium} spin due to an increment of bias voltage. The
accumulated spin in the junction is maintained both by the source
spin and by the bias voltage. If the source spin is turned off, no
spin accumulation is present. The bias voltage is an additional spin
control knob. We consider the case of the source spin in the p
region. As the bias voltage is increased from reverse to forward,
the total accumulated spin in the pn junction decreases, as shown in
Fig. \ref{fig:capacity}. The two main reasons are the decrease of
the built-in field as the bias voltage is increased and the decrease
of the depletion layer. The former inhibits the spin flow, the
latter enlarges the diffusion part of the transport in the p region
during which the spin has more chance to decay before it reaches the
depletion layer.

\section{Spin-polarized pn junctions: bipolar spin-polarized solar cells}

A bipolar solar cell is a pn junction diode which generates oriented
electric current by converting photons into electron-hole pairs in
the active region. The light has to be absorbed within the diffusion
distance of the depletion layer. As the electron-hole pairs are
generated, the minority carriers diffuse towards the depletion layer
in which they are swept to the other side (where they are the
majority carriers). This sweep provides the electric drift and,
eventually, the emf. If the light shines on the depletion layer
directly, electrons are swept to the n, and holes to the p side. A
bipolar electric current flows.

\begin{figure}
\centerline{\psfig{file=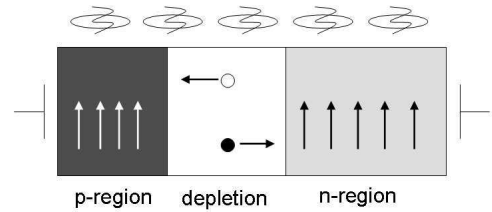,width=0.8\linewidth}}
\caption{Bipolar spin-polarized solar cell. Circularly polarized
light illuminates a pn junction, generating spin-polarized
electron-hole pairs. Electrons (filled circle) in the depletion
layer are swept into the n side, holes (empty circle) into the p
side, by the built-in electric field. The sweep produces the
electromotive force (emf).} \label{fig:solar_cell}
\end{figure}

A bipolar spin-polarized solar cell ~\cite{Zutic2001:APL} is a pn
junction illuminated with a circularly polarized light. See Fig.
\ref{fig:solar_cell}. The light generates both the nonequilibrium
carrier density and spin. In the p region, the light generates
nonequilibrium minority carriers and spin polarization by means of
optical orientation. In the n region, the generated spin
polarization is much lower (generally dependent on the intensity of
light), a result of spin pumping. Since spin-polarized carriers are
generated within the depletion layer, a spin-polarized current flows
which can be used for spin injection. In addition to the density
spin polarization, one is interested here in the spin current
polarization $P_J$, as it is this quantity that determines the spin
injection efficiency. ~\cite{Zutic2004:RMP, Fabian2007:APS}

\begin{figure}
\centerline{\psfig{file=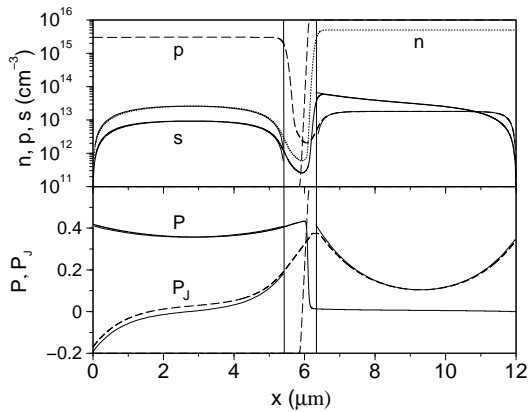,width=0.8\linewidth}}
\caption{Calculated spatial profile of carrier and spin densities
(upper figure) and spin polarizations (lower figure) for a generic
bipolar spin-polarized solar cell under no applied voltage.
Circularly polarized light illuminates the pn junction uniformly,
creating nonequilibrium minority carrier densities ($n$ electrons,
$p$ holes) and spin ($s$) in all regions. The dashed line indicates
the doping profile $N_d-N_a$, the same as in Fig. \ref{fig:density},
while the vertical lines at $x=5.4$ and $x=6.3$ $\mu$m show the
depletion layer boundaries. Some of the curves are accompanied by
model calculations, based on the generalized Shockley model.  From
Ref. \onlinecite{Zutic2001:APL}.} \label{fig:solar_profile}
\end{figure}

The carrier and spin densities, as well as the spin polarization
profiles, are shown in Fig. \ref{fig:solar_profile} for a generic
GaAs inspired pn junction  illuminated uniformly with circularly
polarized light. The electron spin density $s$ follows the electron
density $n$ in the p region. Here the spin orientation is rather
efficient, creating about 40\% spin polarization (the nominal spin
polarization at electron-hole creation is 50\% in zinc-blende
semiconductors). In the n region, in which the density $n$ is
essentially the doping density $N_d$, the spin density decays from
the depletion layer to the far end of the n region, where spin
absorbing boundary conditions are assumed (spin Ohmic contacts). The
spin is pumped into the n region by the light itself, but also by
the minority channel spin pumping. The spin polarization in the n
region remains small. In contrast, the spin current polarization is
actually higher in the n region as compared to the p region. In the
p region the polarization $P_J$ even changes sign, indicating a
negative spin current there. All those behaviors are nicely
explained by the analytical model of the generalized Shockley
conditions; the model calculations are included in Fig.
\ref{fig:solar_profile} as thin lines accompanying the numerical
curves. The negative sign of $P_J$, for example, is given by the
positive gradient of $s$ for $x \alt 2$ $\mu$m. In the minority
region this gradient is the only relevant contribution (that is,
diffusion) to the spin current.

\begin{figure}
\centerline{\psfig{file=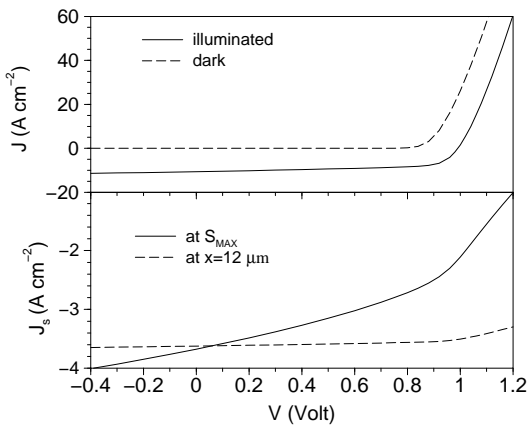,width=0.8\linewidth}} \caption{I-V
characteristics for charge (top) and spin (bottom) currents for a
bipolar spin-polarized solar cell. The charge current is calculated
for an illuminated (solid) and dark (dashed) junction. The spin
current is evaluated at the region of the pn junction in which the
spin density has a maximum $S_{\rm MAX}$ (at $x\approx 6.5$ $\mu$m,
see Fig. \ref{fig:solar_profile}), as well as at the end of the n
region, here $x=12$ $\mu$m. From Ref. \onlinecite{Zutic2001:APL}.}
\label{fig:solar_current}
\end{figure}

Since spin-polarized solar cells are envisioned for spin injection,
the question arises of the I-V characteristic of the spin current.
Selected characteristics are shown in Fig. \ref{fig:solar_current}.
Let us look at the dark charge current first. The I-V curve is the
typical rectification curve: there is a weak generation current at
negative (reverse bias) voltages and the exponentially increasing
current at positive (forward bias) voltages. In the illuminated
diode this current is superimposed on the reverse current due to the
photo emf.

As for the spin current, one needs to select its position in the
diode. For the purposes of illustration, we choose the spin current
at the far end of the diode, here at $x=12$ $\mu$m, which would be
the place of spin injection if the diode were to be used for this
purpose. Another point we choose for calculating the spin current is
the point at which the spin density in the n region has its maximum.
From Fig. \ref{fig:solar_current} we see that the magnitude of the
spin current (note the sign) decreases as the voltage increases.
This is due to the increased importance of spin pumping by the
minority carriers at large reverse biases. The effect is more
pronounced for the point of the maximum spin density, which is
naturally more sensitive to the pumping.

\section{Magnetic pn junction diodes and transistors}

The nonmagnetic spin-polarized devices described above can be
generalized to magnetic bipolar devices, most important being
magnetic pn junction diodes and transistors. These employ magnetic
semiconductors ~\cite{Dietl:2007, Jungwirth2006:RMP}, which are
either ferromagnetic (say, GaMnAs), or semiconductors doped with
magnetic impurities, nominally paramagnetic but with giant g-factors
(on the order of 100) to allow sizable spin Zeeman splitting and
practical equilibrium spin polarization of the current carriers.

Magnetic pn junctions were introduced in Ref.
\onlinecite{Zutic2002:PRL}, while the analytical theory of the
generalized Shockley transport in such devices was laid out in Ref.
\onlinecite{Fabian2002:PRB}. The magnetic diode scheme calls
naturally to be extended to a transistor (the bipolar spin
transistor).~\cite{Fabian2002:P, Fabian2004:APL, Fabian2004:PRB,%
Fabian2004:APP, Fabian2005:APL, Lebedeva2003:JAP, Flatte2003:APL}
The magnetic bipolar spintronics is in detail described in Ref.
\onlinecite{Fabian2007:APS}; we focus on the most relevant physics
here, namely, the spin-charge coupling or the spin-voltaic effect.
As discussed in the next section, this coupling has already been
observed experimentally in magnetic pn junctions.

\begin{figure}
\centerline{\psfig{file=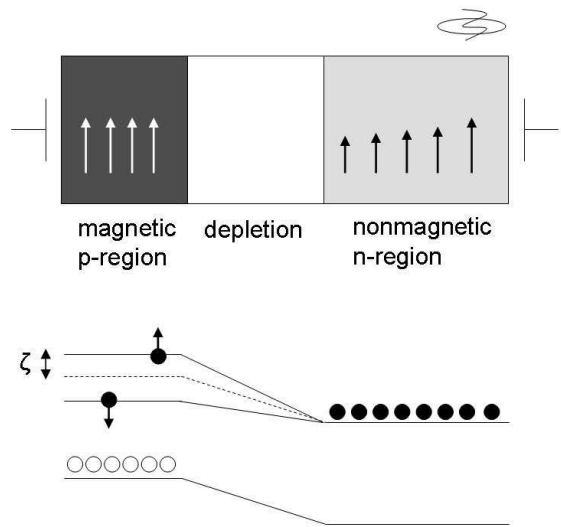,width=1\linewidth}}
\caption{Magnetic pn junction. The p region is magnetic, indicated
by the spin splitting ($2\zeta$) of the conduction band. The n
region is nonmagnetic, but an external spin source, here optical
orientation, generates nonequilibrium spin there.  } \label{fig:md}
\end{figure}

The magnetic diode depicted in Fig. \ref{fig:md} has a magnetic p
region, seen by the spin-split conduction band, and a nonmagnetic n
region. A spin source adds nonequilibrium spin in the n region.
Following the spin-charge coupling idea of Silsbee and Johnson
~\cite{Silsbee1980:BMR, Johnson1985:PRL}, we ask how is the coupling
between the nonequilibrium spin and the equilibrium magnetization
manifested in the I-V characteristics. While the answer is already
given in Eq. \ref{eq:nL}, we reproduce a qualitative argument
presented first in Ref. \onlinecite{Fabian2002:PRB} (inspired by a
similar derivation for conventional diodes in the textbook, Ref.
~\onlinecite{Ashcroft:1976}) to derive the spin-charge effect in the
magnetic pn junction.

Let the electron density at the n side (here $R$ for right) of the
depletion layer be $n_{R\uparrow}$ and $n_{R\downarrow}$ for spin up
and down electrons, respectively. In terms of spin polarization
$P_R$, we have,
\begin{eqnarray}
n_{R\uparrow}& = & n_{R} (1+P_R)/2, \\
n_{R\downarrow}& = & n_{R} (1-P_R)/2.
\end{eqnarray}
The electrons in the n region need to cross a built-in electric
barrier onto the p side. The effective barriers are different for
spin up and down electrons,
\begin{eqnarray}
V_{b\uparrow} & = & V_b + \zeta, \\
V_{b\downarrow} & = & V_b - \zeta,
\end{eqnarray}
where $V_b$ is the intrinsic built-in barrier in the absence of the
equilibrium spin splitting in the p side. In the presence of an
applied bias, both barriers shift to
\begin{eqnarray}
V_{b\uparrow} & = & V_b + \zeta + V, \\
V_{b\downarrow} & = & V_b - \zeta + V.
\end{eqnarray}
Let us calculate the current of electrons flowing from the n to the
p region. Since the electrons need to overcome the barrier, the spin
up and spin down currents (so called recombination, for the label
$r$) will be
\begin{eqnarray}
J_{r\uparrow} & = & K n_{R\uparrow} e^{V_b + \zeta + V}, \\
J_{r\downarrow} & = & K n_{R\downarrow} e^{V_b - \zeta + V}.
\end{eqnarray}
The current is simply proportional to the density of available
electrons and to the Boltzmann thermionic emission factor (we are in
the nondegenerate doping limit) for the probability to find the
electrons of the energy needed to overcome the barrier.

Electrons also flow from the p to the n region. This flow
constitutes the generation current. It is crucial to realize that
this current does not depend on the applied bias $V$. Indeed, the
current is formed by the electrons created by thermal processes
(excitations from a valence into the conduction band) in a proximity
of the depletion layer in which they are swept into the n side. The
current is not limited by the sweep, but by the generation of the
carriers. For the spins in equilibrium as well as at $V=0$, the
individual recombination and generation currents for the given spin
orientation must equal, so that no net current flows. This gives for
the generation currents the following formulas:
\begin{eqnarray}
J_{g\uparrow} & = & K (n_{0R}/2) e^{V_b + \zeta}, \\
J_{g\downarrow} & = & K (n_{0R}/2) e^{V_b - \zeta}.
\end{eqnarray}
The total electron current, $J_n = J_r - J_g$ then is
\begin{equation}
J_n = K n_{0L} \left [ e^V\left ( 1+ P_R P_{0L} \right ) -1 \right
],
\end{equation}
where
\begin{equation}
n_{0L} = n_{0R} e^{-V_b},
\end{equation}
is the equilibrium electron density in the p region for zero spin
splitting there; nondegenerate statistics is used for expressing the
equilibrium spin polarization $P_{0L}$ as a function of $\zeta$.

While the constant $K$ is undetermined by this qualitative argument,
the spin-charge coupling is manifested by the product $\delta P_R
P_{0L}$, which is the product of the nonequilibrium and equilibrium
spin. If the spins are parallel, the electron (and the total)
current is enhanced; vice versa for the antiparallel spins. This
gives rise to the giant magnetoresistance behavior of magnetic pn
junctions. Furthermore, if $V=0$ the current does not vanish! The
reason is that in the nonequilibrium spin converts the spin-charge
coupling into current. We have termed this the spin-voltaic effect
.~\cite{Zutic2002:PRL,Zutic2003:P}

Magnetic pn junctions can also be used to measure
\emph{all-electronically} the spin relaxation time $\tau_s$ in the
nonmagnetic region. The detection of $\tau_s$ would rely on the
sensitivity of the residual, so called spin-voltaic current, as a
function of the bias $V$, on $\tau_s$. The electrically injected (or
optically oriented) spin source is in the nonmagnetic n region. The
spin injection relies on the preservation of the nonequilibrium spin
across the depletion layer. As such, the more nonequilibrium spin
arrives at the depletion layer on the n side, the more it is
injected into the p side. Furthermore, the zero-bias current, which
is proportional to the spin-charge coupling, is proportional to the
nonequilibrium spin at the depletion layer, and thus, to the spin
relaxation time. The more the spin relaxes (in the nonmagnetic
region) the less current there flows. The specifics are discussed in
Ref. \onlinecite{Zutic2003:APL}.

It appears that magnetic semiconductor nanostructures (say, Mn-doped
quantum dots) have desirable materials properties~\cite{Mackowski2004:APL,Leger2006:PRL,Holub2004:APL,%
Govorov2005:PRB, Abolfath2007:Pa, Savic2007:PRB, Cheng2008:PRB} in
that the strong Coulomb interactions together with the quantum
confinement can increase the temperatures at which the magnetization
occurs at much higher values than in the corresponding
bulk~\cite{Fernandez2004:PRL,Holub2004:APL,%
Abolfath2007:PRL} materials.

\section{Experimental realizations}

Two experimental demonstrations of the ideas related to our proposal
for the above bipolar spintronics devices have been reported: a
spin-voltaic effect in a paramagnetic ~\cite{Kondo2006:JJAP} (the
employed photo diode here works as an inverse spin LED) and a
ferromagnetic diode.~\cite{Chen2006:PRB} In the former experiment
the nonequilibrium spin is generated optically in the n region,
giving rise to a spin-voltaic effect via coupling to the equilibrium
paramagnetic Zeeman spin in the p region (analogous to the approach
in Ref. \onlinecite{Dery2006:JAP}). Below we describe the latter
experiment ~\cite{Chen2006:PRB} in more detail.

\begin{figure}
\centerline{\psfig{file=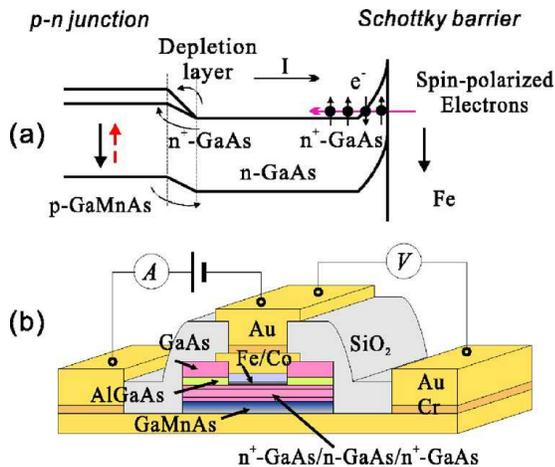,width=1\linewidth}}
\caption{Magnetic GaMnAs/GaAs pn junction diode. The upper figure
shows the electronic bands, with the conduction band spin split in
the ferromagnetic p-GaMnAs. The Schottky barrier serves to inject
spin from the Fe electrode into the n-GaAs region. The lower graph
shows the actual materials and geometry composition of the device,
indicating the four-probe measurement. From Ref.
\onlinecite{Chen2006:PRB}.} \label{fig:chen1}
\end{figure}

A magnetic pn junction is formed by the ferromagnetic GaMnAs, which
is intrinsically p-doped (a significant portion of the 6\% of Mn are
acceptors) and by an n-doped GaAs, which is nominally nonmagnetic
(see Fig. 9). Nonequilibrium source spin is provided by the Fe
electrode, magnetically biased by Co, attached to the n-GaAs region,
with the interface forming a Schottky barrier. Electrons from the Fe
electrode tunnel through the barrier into the n region, bringing
along spin polarization. The injected spin diffuses towards the
depletion layer. There, the spin-charge coupling takes place: if the
nonequilibrium spin is parallel to the equilibrium electron spin in
p-GaMnAs, the current is enhanced. If the spins are antiparallel,
the current is inhibited.

\begin{figure}
\centerline{\psfig{file=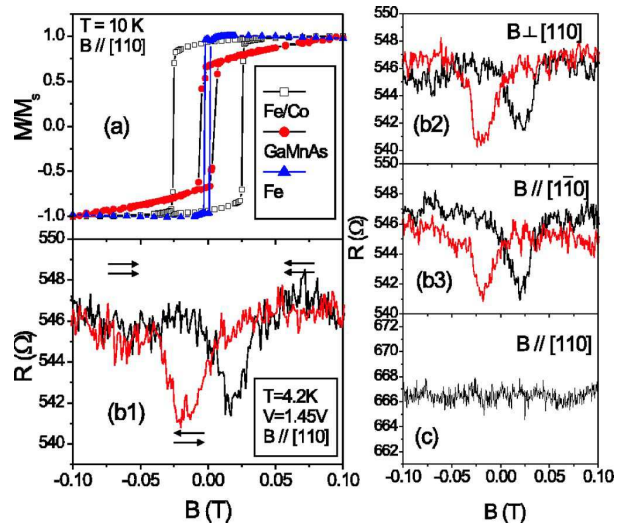,width=1\linewidth}} \caption{(a)
Magnetization hysteresis traces of {Fe/Co}, {GaMnAs}, and Fe films.
The coercive field of the Co biased film is a decade larger than
that of Fe or GaMnAs. (b1 - b3) Magnetoresistances for different
orientations of the applied magnetic fields. Windows with parallel
and antiparallel orientations of the Fe/Co and GaMnAs films are
indicated. (c) Magnetoresistance of the device with magnetically
unbiased Fe film, serving as a reference.  From Ref.
\onlinecite{Chen2006:PRB}.} \label{fig:chen2}
\end{figure}

The spin-charge coupling manifests itself as a spin-voltaic effect,
by producing changes in the resistance for parallel and antiparallel
spin orientations of GaMnAs and the injecting Fe film. The
magnetoresistance traces are plotted in Fig. \ref{fig:chen2}. Since
Fe and GaMnAs films used in the device have similar coercive fields,
the Fe layer was biased by Co, enhancing its coercive field by an
order of magnitude (to about 30 mT). This is seen in Fig.
\ref{fig:chen2} a). With the biased magnetization one should see the
magnetic windows with antiparallel spin orientations, in the
magnetoresistance measurements. Such windows are observed in Fig.
\ref{fig:chen2} b1) through b3), for different orientations of the
applied magnetic field. Finally, a featureless magnetoresistance is
obtained using a magnetically unbiased Fe layer, as seen in Fig.
\ref{fig:chen2} c). The spin-voltaic effect has thus been
demonstrated.

\section{Outlook}

Spin-polarized pn junctions operating under spin orientation
conditions provide a particularly suitable opportunity to
investigate spin transport in different regimes. In addition, the
junctions present perhaps the easiest version of spintronic bipolar
devices offering functionalities such as spin-polarized solar cells,
for generating spin-polarized electrical currents, or spin
capacitors, for control of nonequilibrium spin.

Magnetic diodes and transistors, with added source spin, offer still
new functionalities based on spin-charge coupling. Indeed, magnetic
pn junctions could work as giant magnetoresistive magnetic sensors,
and magnetic transistors as magnetic field controlled current
sources. If magnetic semiconductors are taken as desirable materials
candidates for electronic applications, magnetic bipolar devices
should be considered important contenders for spintronic goals of
controlling electric properties by spin and vice versa.

\acknowledgements

This work was supported by SFB 689, SPP 1285, US ONR, NSF-ECCS
CAREER, CNMS at ORNL, and the CCR at SUNY Buffalo.

\bibliographystyle{apsrev}

\bibliography{../../references_master}

\end{document}